\documentclass[reprint, superscriptaddress, showpacs, amsmath,amssymb,aps, prl, twocolumn]{revtex4-1}

\usepackage{graphicx}
\usepackage{todonotes}
\usepackage{color}
\usepackage[normalem]{ulem}

\begin{document}


\title{Giant Spin Gap and Magnon Localization in the Disordered Heisenberg Antiferromagnet Sr$_2$Ir$_{1-x}$Ru$_x$O$_4$}

\author{Yue Cao}
\email{ycao@bnl.gov}
\affiliation{%
 Condensed Matter Physics and Material Science Department, Brookhaven National Laboratory, Upton, NY 11973
}%

\author{Xuerong Liu}%
\email{xliu@aphy.iphy.ac.cn}
\affiliation{%
 Institute of Physics, Chinese Academy of Sciences, Beijing 100190, China
}%

\author{Wenhu Xu}%
\email{wenhuxu@bnl.govn}
\affiliation{%
 Condensed Matter Physics and Material Science Department, Brookhaven National Laboratory, Upton, NY 11973
}%

\author{Weiguo Yin}%
\affiliation{%
 Condensed Matter Physics and Material Science Department, Brookhaven National Laboratory, Upton, NY 11973
}%

\author{D. Meyers}%
\affiliation{%
 Condensed Matter Physics and Material Science Department, Brookhaven National Laboratory, Upton, NY 11973
}%

\author{Jungho Kim}%
\affiliation{%
 Advanced Photon Source, Argonne National Laboratory, Lemont, IL 60439, U.S.A.
}%

\author{Diego Casa}%
\affiliation{%
 Advanced Photon Source, Argonne National Laboratory, Lemont, IL 60439, U.S.A.
}%

\author{Mary Upton}%
\affiliation{%
 Advanced Photon Source, Argonne National Laboratory, Lemont, IL 60439, U.S.A.
}%

\author{Thomas Gog}%
\affiliation{%
 Advanced Photon Source, Argonne National Laboratory, Lemont, IL 60439, U.S.A.
}%

\author{Tom Berlijn}%
\affiliation{%
 Computer Science and Mathematics Division and Center for Nanophase Materials Sciences, Oak Ridge National Laboratory, Oak Ridge, TN 37831, U. S. A.
}%

\author{Gonzalo Alvarez}%
\affiliation{%
 Computer Science and Mathematics Division and Center for Nanophase Materials Sciences, Oak Ridge National Laboratory, Oak Ridge, TN 37831, U. S. A.
}%

\author{Shujuan Yuan}%
\affiliation{%
 Department of Physics and Astronomy, University of Kentucky, Lexington, KY
}%

\author{Jasminka Terzic}%
\affiliation{%
 Department of Physics and Astronomy, University of Kentucky, Lexington, KY
}%

\author{J. M. Tranquada}%
\affiliation{%
 Condensed Matter Physics and Material Science Department, Brookhaven National Laboratory, Upton, NY 11973
}%

\author{John P. Hill}%
\affiliation{%
 National Synchrotron Light Source II (NSLS-II), Brookhaven National Laboratory, Upton, NY 11973
}%

\author{Gang Cao}%
\affiliation{%
 Department of Physics and Astronomy, University of Kentucky, Lexington, KY
}%
\affiliation{%
 Department of Physics, University of Colorado at Boulder, Boulder, CO 80309
}%

\author{Robert M. Konik}%
\affiliation{%
 Condensed Matter Physics and Material Science Department, Brookhaven National Laboratory, Upton, NY 11973
}%

\author{M. P. M. Dean}%
\email{mdean@bnl.gov}
\affiliation{%
 Condensed Matter Physics and Material Science Department, Brookhaven National Laboratory, Upton, NY 11973
}%

\begin{abstract}
We study the evolution of magnetic excitations in the disordered two-dimensional antiferromagnet Sr$_2$Ir$_{1-x}$Ru$_x$O$_4$. A gigantic magnetic gap greater than 40~meV opens at $x=0.27$ and increases with Ru concentration, rendering the dispersive magnetic excitations in Sr$_2$IrO$_4$ almost momentum-independent. Up to a Ru concentration of $x=0.77$, both experiments and first-principles calculations show the Ir $J_{\text{eff}}=1/2$ state remains intact. The magnetic gap arises from the local interaction anisotropy in the proximity of the Ru disorder. Under the coherent potential approximation, we reproduce the experimental magnetic excitations using the disordered Heisenberg antiferromagnetic model with suppressed next-nearest neighbor ferromagnetic coupling.
\end{abstract}


\def\mathbi#1{\ensuremath{\textbf{\em #1}}}
\def\Q{\ensuremath{\mathbi{Q}}}
\def\SIO{Sr$_2$IrO$_4$}
\def\SIRO{Sr$_2$Ir$_{1-x}$Ru$_x$O$_4$}
\def\SRO{Sr$_2$RuO$_4$}
\def\jone{$\mathit{J}_\textrm{eff}=1/2$}   
\def\jthree{$\mathit{J}_\textrm{eff}=3/2$}
\def\jru{$\mathit{s}=1$}
\def\SRO{Sr$_2$RuO$_4$}

\newcommand{\angstrom}{\mbox{\normalfont\AA}}
\date{\today}

\maketitle

Many of the most interesting phases in condensed matter are accessed by chemically substituting (that is, doping) well-ordered crystalline materials. A particularly notable example is high temperature superconductivity in the cuprates which arises when the quasi-two-dimensional (2D) antiferromagnetic Mott insulating phase in the parent compounds is suppressed. For this reason, understanding the behavior of antiferromagnets in different doping regimes has become a quintessential problem in quantum magnetism \cite{sachdev1999quantum, lee2006doping, vajk2002quantum, scalapino2012common, jia2014persistent}. The majority of experimental work \cite{dean2013persistence,ishii2014high,lee2014asymmetry,dean2015insights} has focused on out-of-plane chemical substitutions that simultaneously introduce mobile carriers and weak disorder \cite{Alloul2009}. In-plane substitutions introduce strong disorder effects and may or may not change the itinerant carrier concentration. Such a situation is less understood \cite{disorder-paper}, and in particular, there is very little information about how magnetic dynamics change upon high doping levels, for example, close to the geometrical percolation threshold $\sim40$\% above which magnetic patches are disconnected \cite{vajk2002quantum}.

The layered iridate \SIO{} has recently emerged as a novel antiferromagnetic insulator with close structural and electronic analogies to the cuprates \cite{kim2008novel,kim2009phase,kim2014fermi,cao2014hallmarks}. Furthermore, single crystals of Sr$_2$Ir$_{1-x}$M$_x$O$_4$ can be produced where \emph{Ir} is substituted with a different transition metal \emph{M} over a wide range \cite{clancy2014dilute,ye2015structure,yuan2015from,calder2015evolution,glamazda2014effects}. This, combined with recent progress in applying resonant inelastic X-ray scattering (RIXS) \cite{ament2011resonant,dean2015insights} to iridates \cite{kim2012magnetic, kim2014excitonic, liu2016anisotropic, gretarsson2016doping, Dean2016} provides an excellent opportunity to determine the behavior of magnetic correlations in disordered (pseudo)spin-$1/2$ Heisenberg antiferromagnets.

In this \emph{Letter}, we investigate the evolution of magnetic excitations in the heavily disordered regime of \SIRO{}. The magnetic correlations survive at least to a high Ru doping of $x=0.77$. A magnetic gap greater than 40~meV develops as early as $x=0.27$, and increases with higher Ru dopings. Eventually the magnetic excitations become localized, and non-dispersive throughout the Brillouin zone. We present a quantitative description of the observed antiferromagnetic excitations using density functional theory and coherent potential approximation (CPA). The giant magnetic gap and the flattened magnetic excitations originate from local orbital anisotropy, as well as the suppression of next-nearest neighbor magnetic interactions.




\begin{figure}
\includegraphics[width=\columnwidth]{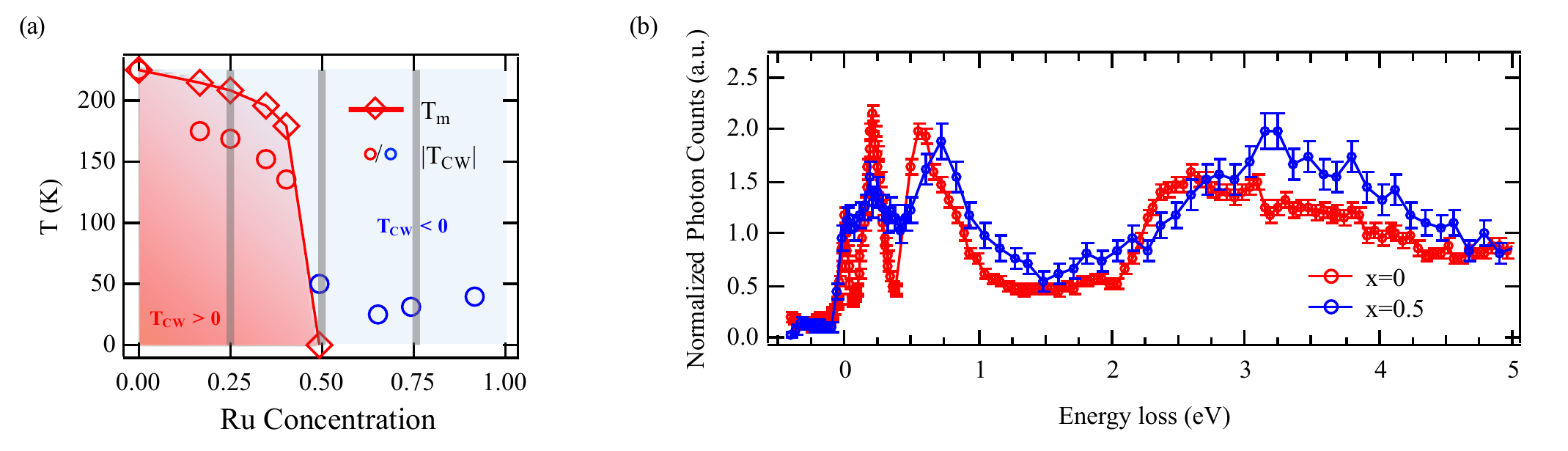} %
\caption{(Color online) (a) The magnetic phase diagram of \SIRO{}. $T_\text{m}$ is the onset temperature for an increased magnetization under an external field of 0.1~T. Around $x=0.50$, $T_\text{m}$ is suppressed to zero. This is also where the Curie-Weiss temperature, $T_\text{C-W}$, changes sign \cite{yuan2015from}. Grey bars mark the select Ru concentrations studied in this work. (b) RIXS spectra of \SIO{} and Sr$_2$Ir$_{0.5}$Ru$_{0.5}$O$_4$ at $(\pi,0)$ over a 5.5~eV window, showing persistent excitations centered around 200~meV, 700~meV and 3.5~eV. These are assigned to magnetic, intra-$t_{2g}$ and $t_{2g} \rightarrow e_g$ processes respectively.} 
\label{Fig1}
\end{figure}

Single crystals of \SIRO{} were grown from off-stoichiometric quantities of SrCl$_2$, SrCO$_3$, IrO$_2$, and RuO$_2$ using self-flux techniques \cite{yuan2015from}. Fig.~\ref{Fig1} (a) shows the magnetic behavior of \SIRO{} which undergoes a cross-over from antiferromagnetic (AFM) to paramagnetic (PM) around \emph{x}~$\sim$~0.50. Further information on sample growth and characterizations are provided in Ref.~\cite{yuan2015from} and the Supplementary Information \cite{som}.


The RIXS measurements were performed at the Ir $L_3$-edge using the MERIX endstation (27-ID-B) at the Advanced Photon Source, Argonne National Laboratory, with a total energy resolution of 80~meV (full width at half maximum) and momentum resolution of 0.23\AA$^{-1}$. All data presented were taken at the base temperature of the cryostat $\sim$12~K. Figure~\ref{Fig1} (b) plots RIXS spectra of \SIRO{} over a wide energy window up to 5~eV. There are three distinctive energy-loss features, around ~200~meV, 700~meV, and 3.5~eV, respectively. The latter two peaks are orbital excitations of the hole in the valence band, providing information on the electronic configuration of the doped and parent compounds. Specifically, the 3.5~eV peak corresponds to the $t_{2g}\rightarrow e_{g}$ excitation. Its increase in energy with Ru concentration is consistent with what is expected due to structural changes. Going from $x=0$ to $x=0.50$ doping, the Ir-O octahedra elongate by 1.0\% along the apex, and contract by 2.9\% in-plane \cite{yuan2015from}, increasing the crystal field splitting, and moving the $t_{2g}\rightarrow e_{g}$ feature to higher energies. By comparing to \SIO{}, we assign the energy-loss peak around 700~meV to the intra-$t_{2g}$ transition, or more precisely, the transition between the Ir $5d$ \jone{} and \jthree{} states \cite{kim2012magnetic,kim2014excitonic}. By the same token, we deem the energy-loss peak around ~200~meV as magnetic in nature, arising from the pseudospin flip.

With increasing Ru concentration all three peaks persist with comparable energy-scales. Thus the Ir electronic configuration does not change dramatically, and the \jone{} and \jthree{} states are still present at these higher Ru concentrations. In \SIO{} and \SRO{}, Ir$^{4+}$ and Ru$^{4+}$ have formal electron configurations of $5d^5$ and $4d^4$, respectively. As dc-resistivity reduces with Ru concentration \cite{yuan2015from}, one might na\"{i}vely assume that the numbers of $d$-electrons on Ru and Ir get closer to an average of 4.5, due to increased electron itinerancy. This would indicate effective hole doping on the Ir site into the \jone{} level, with drastic changes in the local electronic structure. In contrast, our data seem to suggest that Ir maintains a formal valence of $4^+$, which we will now examine using first-principles calculations.

We calculate the orbital configuration of \SIRO{} in the GGA+$U$ implementation of the density functional theory \cite{som}. Table~\ref{d-el-num} lists the calculated electron occupation number on the Ir and Ru $d$-orbitals vs.\ Ru concentration. Up to $x>0.75$ the numbers of $d$-electrons on Ir/Ru have relatively small changes compared to the doped Ru concentration, especially when a spin-polarized ground state is considered. Thus both Ir/Ru sites maintain a valence close to $4^+$, similar to those in \SIO{}/\SRO{}, in agreement with the X-ray absorption experiments in Ref.~\cite{calder2015evolution}. We project the $d$ electron density of states onto the Ir/Ru orbitals, and find that the \jone{} and \jthree{} are indeed robust. This is because for each individual Ir-O octahedron, the spin-orbit coupling energy scale \cite{kim2008novel,liu2012testing} ($\sim400$~meV) dominates over the tetragonal splitting between $t_{2g}$ levels, giving rise to a relatively well-defined pseudospin-1/2 state \cite{jackeli2009mott,liu2012testing}. This interpretation also agrees with the persistence of the insulating phase up to $x\simeq0.50$ in \SIRO{} \cite{yuan2015from}. In our calculation, the electrons near the Fermi level come primarily from the local Ru-O octahedra, and cannot move freely especially for low Ru concentrations.  GGA+$U$ calculations for all Ru concentrations favor spin-polarized ground-states, with antiferromagnetic couplings between the nearest-neighbor Ir \jone{} and Ru \jru{} (pseudo-)spins. We therefore consider that the effect of Ru doping in \SIRO{} is primarily to introduce substitutional \jru{} magnetic disorder (rather than charge doping) for a large range of Ru concentration (until the material gets sufficiently close to \SRO{}). We show later that phenomenological CPA simulations, based on this picture, account for the observed magnetic dispersion.

\begin{table}
\begin{center}
\begin{tabular*}{0.4\textwidth}{@{\extracolsep{\fill} } r | c c c c}
    \hline\hline
    Ru concentration & 0\% & 25\% & 50\% & 100\% \\ \hline
    Ir \qquad NM & 5.20 & 5.16 & 5.19 &  \\
    SP & 5.20 & 5.19 & 5.21 &  \\ \hline
    Ru \qquad  NM &  & 4.53 & 4.61 & 4.47 \\
    SP &  & 4.58 & 4.47 & 4.46 \\
    \hline\hline
\end{tabular*}
\caption{The number of d electrons on each Ir/Ru site, derived from GGA+U calculations. We used $U = 2$~eV \cite{kim2008novel} for Ir and $U = 3$~eV for Ru. NM and SP stand for the non-magnetic and spin-polarized calculations respectively.}\label{d-el-num}
\end{center}
\end{table}


\begin{figure*}
\includegraphics[width=0.8\textwidth]{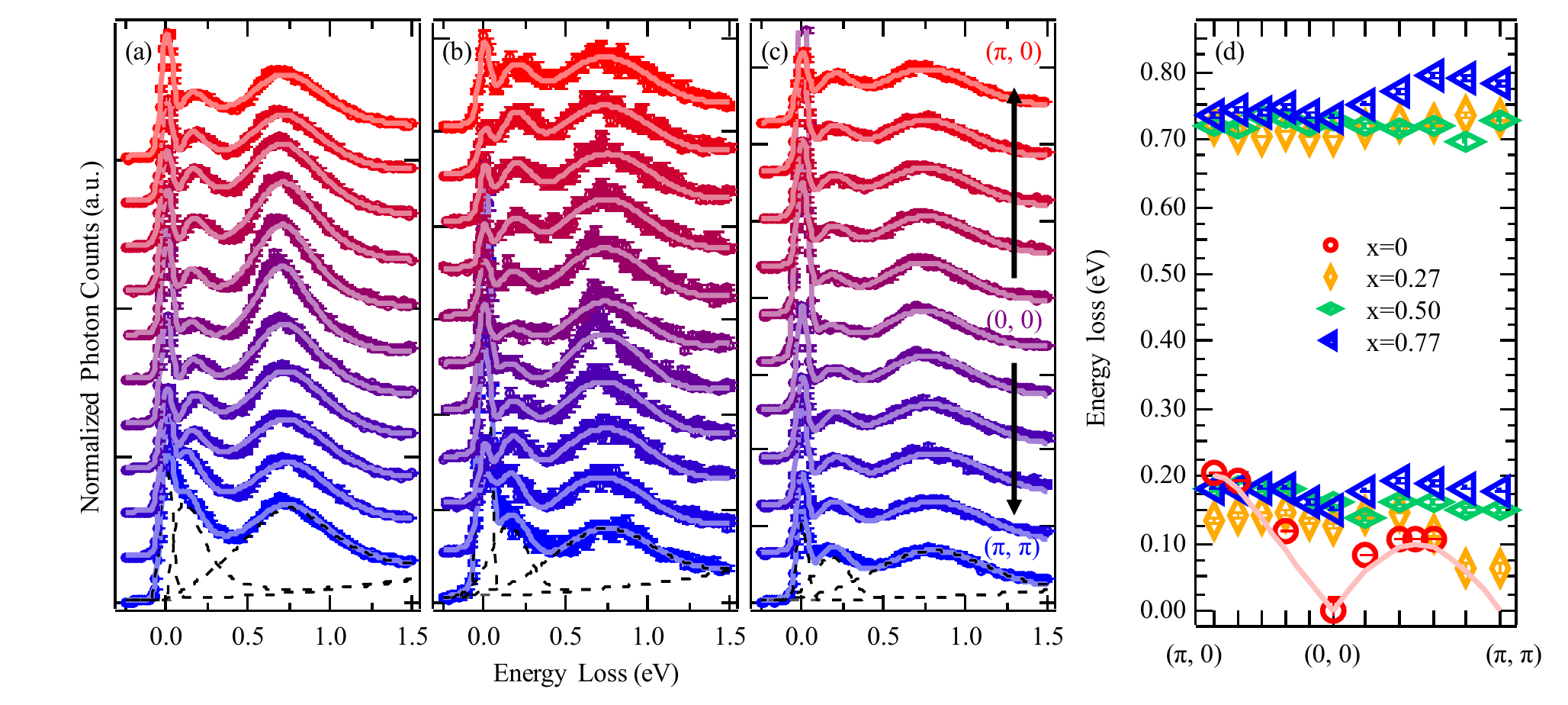} %
\caption{(Color online) (a)-(c) The energy-loss spectra along $(\pi, 0)$--$(0, 0)$--$(\pi, \pi)$ high symmetry directions for three Ru dopings (a) $x=0.27$, (b) $x=0.50$, and (c) $x=0.77$. The solid thick lines are fits to the RIXS data. The dashed black curves are the individual components of the fit for the energy loss curve at $(\pi, \pi)$. (d) Energies of the magnetic and orbital excitations as a function of momentum transfer for different Ru concentrations.  Error bars represent the uncertainty from the least-mean square fitting algorithm. The pink solid line is the fitted magnon dispersion in \SIO{} reproduced from Ref.~\cite{kim2012magnetic}.}
\label{Fig2}
\end{figure*}

In Fig. \ref{Fig2}(a)-(c) we take a closer look at the excitations within the first 1.5~eV of energy loss as \SIRO{} crosses the AFM-PM phase boundary: (a) $x=0.27$, (b) $x=0.50$, and (c) $x=0.77$. All dopings are marked with grey bars in the phase diagram in Fig.~\ref{Fig1} (a). We assign the $(\pi, 0)$ direction to be parallel to the nearest neighbor Ir-Ir bond directions, in analogy with the usual definition in square-net cuprates. As expected in a disordered system, both the magnetic and orbital excitations are broader than in the undoped \SIO{}. We fit the elastic peak and the two excitations with three Gaussian peaks on top of a smooth background, and the fitted energies of the magnetic and orbital modes are shown in Fig.~\ref{Fig2} (d).

The most striking feature is that the antiferromagnetic excitations persist up to at least $x=0.77$, and that the maximum energy scale of the magnetic excitations at high dopings is comparable to that in undoped \SIO{}. In a simple mean-field description, the overall magnetic excitation energy would be expected to decrease appreciably due to lower magnon energies in \SRO{}. The results here reflect the strong local correlation in the material that voids mean-field descriptions. Such a result is reminiscent of studies of electron and hole doped cuprates and iridates \cite{vignolle2007two, lipscombe2007persistence, dean2012spin, dean2013persistence,lee2014asymmetry,ishii2014high,dean2015insights, liu2016anisotropic,gretarsson2016doping}, albeit up to smaller maximum doping level of 40\% in cuprates and a mere 10\% in iridates. Specifically, the magnetic excitation energies at ($\pi, 0$) are robust against doped charge carriers and in this work disorder, while profound changes take place around e.g. ($\pi$, $\pi$) and ($\pi/2$, $\pi/2$) \cite{lee2014asymmetry,ishii2014high,liu2016anisotropic,gretarsson2016doping,fujita2011progress}. This similarity is not necessarily expected as it involves comparing the effects of itinerant carriers introduced by out-of-plane atomic substitutions with in-plane replacement of the Ir atoms.

The second observation is that a large spin gap already opens for $x=0.27$, and appears to increase and saturate with Ru doping across the phase transition. This is in sharp contrast to the dispersive, almost gapless paramagnetic excitations in the electron and hole doped cuprates and iridates mentioned above. At higher doping of $x=0.50$ and $x=0.77$, the magnetic modes are almost dispersionless within an energy range of 150-180~meV. Such an energy scale lies between the zone boundary energy-scales of \SIO{}: $\sim 200$~meV at $(\pi, 0)$ and $\sim100$~meV at $(\pi/2, \pi/2)$.


%


\begin{figure}
\includegraphics[width=0.6\columnwidth]{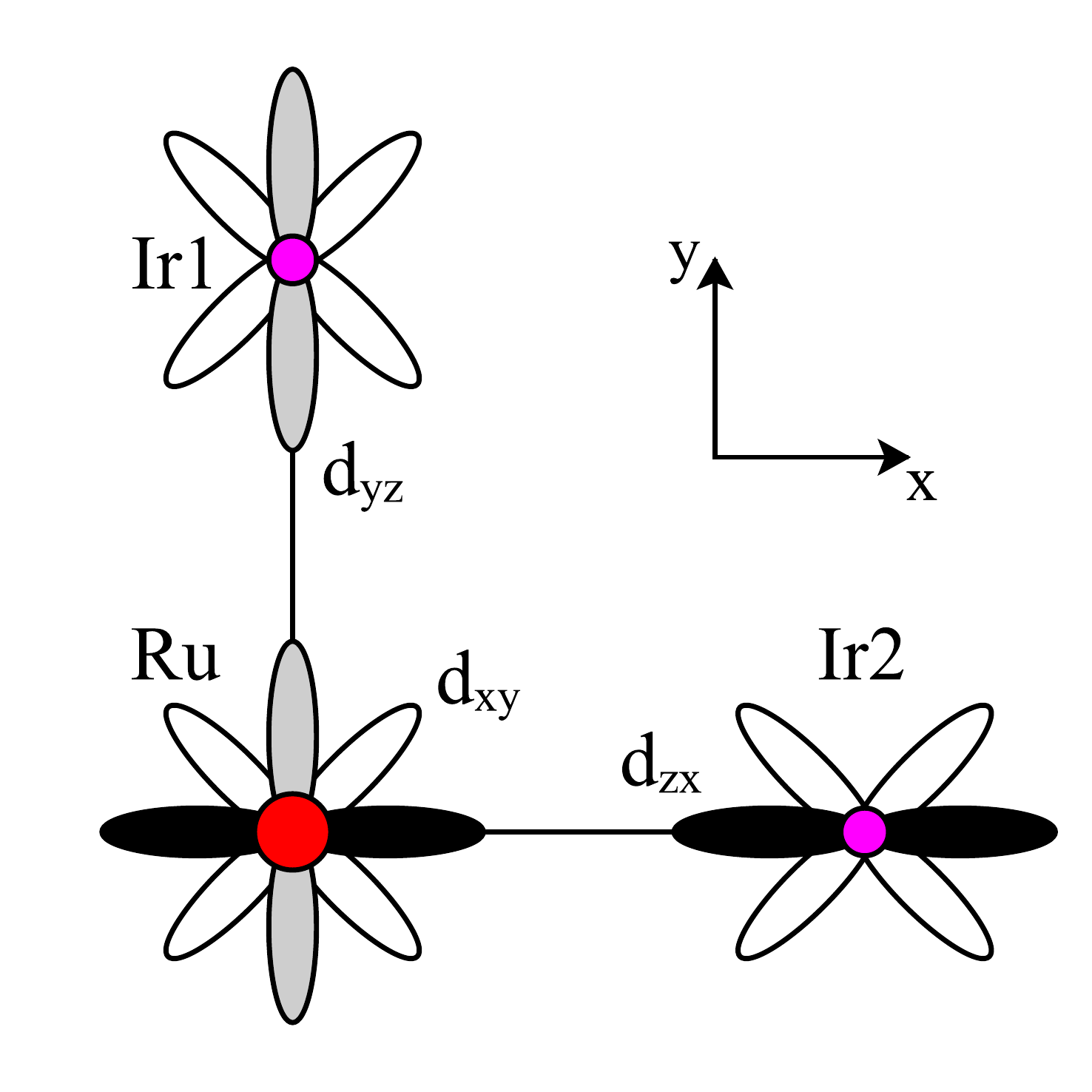} 
\caption{ (Color online) Geometry of Ru-Ir bonds with orbitals active $t_{2g}$ orbitals along these bonds (top view). Black: $d_{zx}$ orbitals. Grey: $d_{yz}$ orbitals. White: $d_{xy}$ orbitals. The oxygen atoms between the nearest-neighbor Ir atoms are not displayed, and the oxygen states have been projected out in the effective superexchange model.}
\label{t2g_exchange}
\end{figure}

The opening up of a finite gap in the spin excitation spectrum can be explained by broken continuous rotational symmetry. In our case, a plausible origin of spin gap is the anisotropic antiferromagnetic exchange between $t_{2g}$ electrons experiencing different spin-orbit coupling on the Ru $4d$ and Ir $5d$ orbitals \cite{PhysRevLett.111.057202}, as well as the strong Hund's coupling on the Ru $4d$ orbitals. We illustrate this by considering the toy model in Fig.~\ref{t2g_exchange}. For a pair of Ru-Ir spins, after projecting out the oxygen states, the effective magnetic Hamiltonian $\mathcal{H} = \mathcal{H}_\text{SO} + \mathcal{H}_\text{Hund} + \mathcal{H}_\text{AF}$ consists of the spin-orbit coupling $\mathcal{H}_\text{SO}$ on the Ir sites, the Hund's coupling $\mathcal{H}_\text{Hund}$ between the $t_{2g}$ spins on the Ru site, and the antiferromagnetic coupling between Ru and Ir spins, $\mathcal{H}_\text{AF} = J_{yz} \mathbf{s}_{\text{Ru}, yz} \cdot \mathbf{s}_{\text{Ir1}, yz} + J_{xy} \mathbf{s}_{\text{Ru}, xy} \cdot \mathbf{s}_{\text{Ir1}, xy} + J_{zx}\mathbf{s}_{\text{Ru}, zx} \cdot \mathbf{s}_{\text{Ir2}, zx} + J_{xy} \mathbf{s}_{\text{Ru}, xy} \cdot \mathbf{s}_{\text{Ir2}, xy}$. As a first-order perturbation to $\mathcal{H}_\text{SO}$, $\mathcal{H}_\text{AF}$ induces an effective anisotropic coupling between the $t_{2g}$ spins on Ru and the pseudospin (instead of physical spin) $\mathbf{J}_\text{eff}$ on Ir~\cite{PhysRevLett.111.057202}. For example, the effective coupling on the Ru-Ir2 bond, $\mathcal{H}_{AF2} = J_{zx}/3 (s_{\text{Ru}, zx}^{x} J_\text{Ir2, eff}^{x} - s_{\text{Ru}, zx}^{y} J_\text{Ir2, eff}^{y} + s_{\text{Ru}, zx}^{z} J_\text{Ir2, eff}^{z}) + J_{xy}/3 (s_{\text{Ru}, xy}^{x} J_\text{Ir2, eff}^{x} + s_{\text{Ru}, xy}^{y} J_\text{Ir2, eff}^{y} + s_{\text{Ru}, xy}^{z} J_\text{Ir2, eff}^{z})$, has an easy $z-x$ plane. Similarly, the effective coupling on the Ru-Ir1 bond has an easy $y-z$ plane. The Hund's coupling on the Ru site, $\mathcal{H}_{\text{Hund}} = -J_{H} \mathbf{S}_{\text{Ru}}^2$, where $\mathbf{S}_{\text{Ru}} = \mathbf{s}_{\text{Ru}, xy} + \mathbf{s}_{\text{Ru}, yz} + \mathbf{s}_{\text{Ru}, xz}$ favors aligning the Ru $t_{2g}$ spins. If one Ru site is connected to neighboring Ir sites by at least two perpendicular Ru-Ir bonds (as in Fig.~\ref{t2g_exchange}), the degeneracy of rotating $\mathbf{s}_{\text{Ru},yz}$ in the $y-z$ plane and $\mathbf{s}_{\text{Ru},zx}$ in the $z-x$ plane will be lifted by $\mathcal{H}_{\text{Hund}}$, giving rise to the effective anisotropic exchange with an easy $z$-axis.

\begin{figure}
\includegraphics[width=0.8\columnwidth]{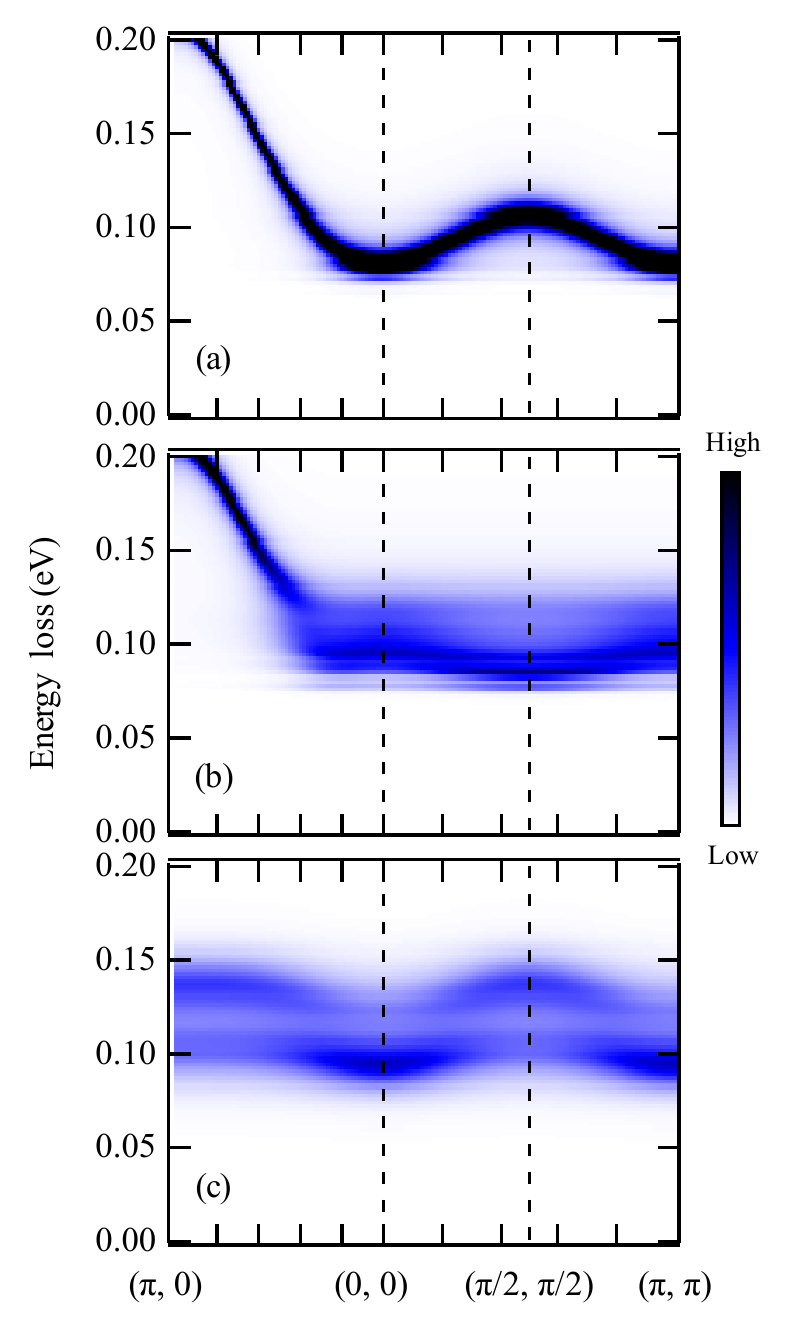} %
\caption{(Color online) (a) and (b) Magnetic excitations calculated using CPA with (a) $x=0.25$ and (b) $x=0.50$ concentrations of \jru{} moments (to simulate the Ru sites), preserving both the nearest neighbor (NN) and next-nearest neighbor (NNN) magnetic interactions between \jone{} sites. (c) Calculated magnetic excitations with 50\% \jru{} moments. The couplings between the NNN and further \jone{} moments are suppressed.}
\label{Fig4}
\end{figure}

We capture the geometrical impact of introducing spin disorder and spin anisotropy on the system using the coherent potential approximation (CPA)~\cite{buyers1971disordered, buyers1972CPA} following the recipe in Ref.~\cite{buyers1972CPA}, and calculate the magnetic excitation spectrum (Fig.~\ref{Fig4}) \cite{CPA_details}. The nearest neighbor (NN) coupling between Ir sites is selected to be $J_{Ir-Ir}$ = 60~meV, identical to that in \SIO{}. We extract nearest-neighbor Ir-Ru coupling $J_{Ir-Ru}\simeq$60~meV, from the energy differences between the magnetic and paramagnetic groundstates calculated using GGA+$U$.  Other input exchange energies for the CPA calculation includes the exchange between next (NNN) and next-next nearest (NNNN) neighboring \jone{} pseudospins,  $J'_{Ir-Ir} = -20$~meV, and $J''_{Ir-Ir} = 15$~meV, respectively~\cite{kim2012magnetic,kim2014excitonic}. As to the exchange between NN \jru{} spins, there is no qualitative change in the simulated magnetic excitation for a weak antiferromagnetic $J_\text{Ru-Ru} \lesssim 5$~meV. Fig.~\ref{Fig4} (a) and (b) show increased magnetic gaps at $(0, 0)$ and $(\pi, \pi)$ with rising \jru{} concentrations. At $x=0.50$ in Fig.~\ref{Fig4} (b), the magnetic excitations are damped and flattened around much of the Brillouin zone, except along $(\pi,0)$---$(0,0)$ where the excitation spectrum is barely affected by the disorder. To further increase the simulated magnetic gap, we set $J'_{Ir-Ir}$ and $J''_{Ir-Ir}$ to zero, and the calculations are shown in Fig.~\ref{Fig4} (c). The magnetic excitation is turned into a heavily damped localized mode, recapturing the dispersionless feature in Fig.~\ref{Fig2} (d). Our suppression of the NNN and NNNN exchange energies may reflect that the corresponding magnetic couplings are destroyed geometrically in the presence of more \jru{} disorders.

The magnetic excitation energies at the two centers of the antiferromagnetic Brillouin zone $(0, 0)$ and $(\pi, \pi)$ are not identical in the $x=0.27$ sample, unlike those in the \SIO{} (Fig.~\ref{Fig2} (d), also see Ref.~\cite{kim2012magnetic,kim2014excitonic}). The observed inequivalence may signal a breakdown of the antiferromagnetic Brillouin zone, which is well-defined \emph{only} under translation symmetry. Admittedly, this effect is not captured in the current CPA calculation. However, in the carrier-doped Mott insulators, where the antiferromagnetic Brillouin zone is also ill-defined, there are theoretical proposals showing different magnetic excitation energies at $(0, 0)$ and $(\pi, \pi)$. For example, using more rigorous determinant quantum Monte Carlo methods, Jia \emph{et al.} \cite{jia2014persistent} interpreted the meltdown of the antiferromagnetic Brillouin zone as associated to the change in spin correlation from antiferromagnetic to ferromagnetic. Such an effect can also be shown in the Schwinger boson treatment of the square-lattice quantum antiferromagnet using the ferromagnetic configuration instead of the N\'{e}el state as the reference state, in which the long-range antiferromagnetic order is generated via Bose-Einstein condensation at $(\pi, \pi)$ with a considerable spin gap at $(0, 0)$ \cite{yin1996dynamics}. It is to be noted that the antiferromagnetic and/or structural Brillouin zone also fails to describe the complete electronic structures in (effectively) carrier doped \SIO{}, as observed in angle-resolved photoemission experiments \cite{cao2014hallmarks,de2015collapse,kim2016observation}.


In summary, we have measured the magnetic excitations in \SIRO{} across a wide doping range and observed a cross-over from dispersive to gapped, localized magnons. First-principles calculations and simulations using the coherent potential approximation provide a thorough description of these findings based on disorder and local anisotropy effects. Similar effects are likely to be at play in other heavily doped transition metal oxides with important implications for understanding the importance of disorder on magnetic correlations and how this might relate to emergent phenomena such as high-Tc superconductivity.



\begin{acknowledgements}
The authors acknowledge fruitful discussions with Gilberto Fabbris and Daniel Haskel. The work at Brookhaven National Laboratory was supported by the U. S. Department of Energy, Division of Materials Science, under Contract No. DE-SC0012704. X.R.L.\ receives financial support from MOST (No. 2015CB921302), CAS (Grant No: XDB07020200), and by the National Thousand Young-Talents Program of China. T.~B. and G.~A. received funding from the Center for Nanophase Materials Sciences, sponsored by the Scientific User Facilities Division, Basic Energy Sciences, Department of Energy (DOE), USA, under contract with UT-Battelle. The work at the University of Kentucky was supported by NSF through Grant DMR-1265162. This research used Sector 27 of the Advanced Photon Source, a U.S. Department of Energy (DOE) Office of Science User Facility operated for the DOE Office of Science by Argonne National Laboratory under Contract No. DE-AC02-06CH11357 and beamline X22C of the National Synchrotron Light Source, a U.S. Department of Energy (DOE) Office of Science User Facility operated for the DOE Office of Science by Brookhaven National Laboratory under Contract No. DE-AC02-98CH10886.
\end{acknowledgements}


\begin{thebibliography}{40}
\bibitem [{CPA()}]{CPA_details}%
  \BibitemOpen
  \href@noop {} {}\bibinfo {note} {Sophisticated direct simulations of disorder
  with many large supercells will be published elsewhere.}\BibitemShut {Stop}%
\end{thebibliography}

%

\end{document}